\def\g{gamma-ray }
\def\Jninth{PSR~J0034$-$0534 }
\def\JninthN{PSR~J0034$-$0534}
\def\JninthC{PSR~J0034$-$0534, }

\def\Jfirst{PSR~J1939+2134 }

\def\JfirstC{PSR~J1939+2134, }

\def\Jblack{PSR~J1959+2048 }

\def\JblackC{PSR~J1959+2048, }
\def\JblackF{PSR~J1959+2048. }
\def\JRansom{PSR~J2214+3000 }
\def\JRansomC{PSR~J2214+3000, }
\def\gaeq{\raise.2ex\hbox{$>$}\kern-.75em\lower.9ex\hbox{$\sim$}\,}
\def\laeq{\raise.2ex\hbox{$<$}\kern-.75em\lower.9ex\hbox{$\sim$}\,}

\documentclass[twocolumn,twoside,slac_two]{revtex4}
\usepackage{graphicx}
\usepackage{fancyhdr}
\begin{document}

\title{Modeling Light Curves of the Phase-Aligned Gamma-ray Millisecond Pulsar Subclass}

%

\author{C. Venter}
\affiliation{Centre for Space Research, North-West University, Potchefstroom Campus, Private Bag X6001, Potchefstroom 2520, South Africa}
\author{T.~J. Johnson}
\affiliation{Astrophysics Science Division, NASA Goddard Space Flight Center, Greenbelt, MD 20771, USA\\
Department of Physics, University of Maryland, College Park, MD 20742, USA\\
National Research Council Research Associate, National Academy of Sciences, Washington, DC 20001, resident at Naval Research Laboratory, Washington, DC 20375, USA
}

\author{A.~K. Harding}
\affiliation{Astrophysics Science Division, NASA Goddard Space Flight Center, Greenbelt, MD 20771, USA}

\begin{abstract}
The \g population of millisecond pulsars (MSPs) detected by the \textit{Fermi} Large Area Telescope (LAT) has been steadily increasing. A number of the more recent detections, including \JninthC \Jfirst (B1937+21; the first MSP ever discovered), \Jblack (B1957+20; the first black widow system), and \JRansomC exhibit an unusual phenomenon: nearly phase-aligned radio and gamma-ray light curves (LCs). To account for the phase alignment, we explore geometric models where both the radio and gamma-ray emission originate either in the outer magnetosphere near the light cylinder ($R_{\rm LC}$) or near the polar caps (PCs). We obtain reasonable fits for the first three of these MSPs in the context of ``altitude-limited'' outer gap (alOG) and two-pole caustic (alTPC) geometries. The outer magnetosphere phase-aligned models differ from the standard outer gap (OG)~/ two-pole caustic (TPC) models in two respects: first, the radio emission originates in caustics at relatively high altitudes compared to the usual low-altitude conal radio beams; second, we allow the maximum altitude of the gamma-ray emission region as well as both the minimum and maximum altitudes of the radio emission region to vary within a limited range. Alternatively, there also exist phase-aligned LC solutions for emission originating near the stellar surface in a slot gap (SG) scenario (``low-altitude slot gap'' (laSG) models). We find best-fit LCs using a Markov chain Monte Carlo (MCMC) maximum likelihood approach~\cite{Johnson11}. Our fits imply that the phase-aligned LCs are likely of caustic origin, produced in the outer magnetosphere, and that the radio emission may come from close to $R_{\rm LC}$. We lastly constrain the emission altitudes with typical uncertainties of $\sim0.3R_{\rm LC}$. Our results describe a third gamma-ray MSP subclass, in addition to the two (with non-aligned LCs) previously found~\cite{Venter_MSP09}: those with LCs fit by standard OG / TPC models, and those with LCs fit by pair-starved polar cap (PSPC) models.
\end{abstract}

\maketitle

\thispagestyle{fancy}


\section{INTRODUCTION}
\label{sec:intro}
The first pulsar catalog released by \textit{Fermi} Large Area Telescope (LAT) included~46 \g pulsars \cite{Abdo09_Cat}, 8 of which were millisecond pulsars (MSPs)~\cite{Abdo09_MSP}. Currently, there are $>20$ \g MSPs~\cite{Guillemot11} and $>70$ \g pulsars in total~\cite{Romani11}. 
The discovery of \Jninth \cite{Abdo09_J0034} revealed it to be the first MSP to have (nearly) phase-aligned radio and \g light curves (LCs). This rare phenomenon has only been observed for the Crab pulsar~\cite{Kniffen74}. However, this behavior has now also been observed for \Jfirst (B1937+21), \Jblack (B1957+20)~\cite{Guillemot11_2msps}, and \JRansom \cite{Ransom11}, and more MSPs will be added to this subclass. 

\subsection{Traditional Emission Models}
Two classes of pulsar models have been used to describe high-energy (HE) pulsar emission. In polar cap (PC) models~\cite{Daugherty82,DH96}, primary electrons are ejected from the neutron star (NS) surface and accelerated along curved magnetic field lines, producing curvature radiation gamma rays. Thermal X-rays may also be upscattered to \g energies. Subsequently, these gamma rays are converted into electron-positron pairs via magnetic pair production in the intense magnetic fields close to the stellar surface (at radius $R_{\rm NS}$). In addition,
a slot gap (SG) \cite{Arons83,MH03_SG} may form along the last open magnetic field lines of the pulsar magnetosphere in the absence of pair creation along those lines. This corresponds to a two-pole caustic (TPC) geometry \cite{Dyks03} which may extend from the stellar surface up to near the light cylinder (at radius $R_{\rm LC}$). Outer gap (OG) models \cite{CHR86a,Romani96} represent the second model class. In these models, HE radiation is produced along the last open field lines above the null charge surface (NCS) where the Goldreich-Julian charge density changes sign. 
The narrow gaps in both the OG and TPC models require screening of the electric field parallel to the local magnetic field, and therefore presupposes copious pair production. Lastly, HE LCs were also modeled in the context of OG and TPC models in a force-free magnetic field geometry, proposing a separatrix layer model close to $R_{\rm LC}$~\cite{BS10}.

\subsection{Formation of Caustics}
\label{sec:caustics}
HE photons escaping from the magnetosphere are subject to two relativistic effects: their traveling direction is aberrated due to the large corotation velocity, while their arrival time at the observer is determined by their emission height, due to the finite speed of light. Lastly, it is assumed that these photons are emitted tangent to the local magnetic field lines in the co-rotating frame. The combination of these three effects result in the formation of caustics, i.e., the accumulation of photons in narrow phase bands~\cite{Morini83}. These caustics manifest themselves as bright peaks in the observed pulse profiles. 

\subsection{MSP Models}
Due to their much lower surface dipole magnetic field strengths, MSPs have been thought to have pair-starved magnetospheres, where the magnetic pair multiplicity is not high enough to screen the accelerating electric field in the open volume above the PC~\cite{HUM05,VdeJ05}. In this case, primary electrons are accelerated up to very high altitudes above the full PC, while pair formation is suppressed. This model is called the pair-starved polar cap (PSPC) model~\cite{MH04_PS,MH09_PS}, an extension of the traditional PC model. MSP LCs and spectra have been modeled using this framework~\cite{Frackowiak05,HUM05,VdeJ05}.
Alternatively, MSP spectra and energetics have also been modeled in the context of an OG model \cite{Zhang03,Zhang07}. An annular gap model~\cite{Du10} can furthermore reproduce the main characteristics of the \g LCs of three MSPs, although this model does not attempt to model the nonzero phase offsets between the \g and radio profiles. 

\subsection{MSP Subclasses}
The first~8 \textit{Fermi}-detected \g MSPs have been modeled~\cite{Venter_MSP09}. Two distinct MSP subclasses were found: those whose LCs are well fit by a standard OG or TPC model, and those whose LCs are well fit by a PSPC model (with these fits being mutually exclusive). These models yielded the correct radio-to-gamma phase lags when the radio emission was modeled as a cone beam at lower altitude. Such fits implied that MSPs have screened magnetospheres with large amounts of pairs available, as these conditions are needed to set up the gap structure presupposed by the OG / TPC models. Small distortions of the dipole magnetic field causing offsets of the PC may provide a mechanism for enhancing pair creation, even in low-spin-down pulsars~\cite{HM11}. 
This paper discusses a third sublcass of MSPs: those having phase-aligned radio and \g LCs.

\subsection{Motivation for Caustic Radio Emission}
In contrast to the first two MSP sublcasses, the near phase-alignment of the \g and radio LCs of MSPs in the third subclass argues for overlapping emission regions. These co-located emission regions may occur at high altitudes, so that the radio emission will be subject to the same relativistic effects as the \g emission, as described in Section~\ref{sec:caustics}. Discoveries of new \g MSPs exhibiting phase-aligned LCs therefore motivate the investigation of high-altitude~\cite{Manchester05} caustic radio emission. 

A second argument motivating caustic radio emission comes from investigating the beaming properties of normal pulsars and MSPs detectable using blind searches on \g data as well as radio data~\cite{Ravi10}. The relative number of \g to radio pulsars for each of these `gamma-ray-selected' and `radio-selected' samples implies that radio and \g beams must have comparable sky coverage of $\sim4\pi$~sr for pulsars with high spin-down luminosities ($\dot{E}_{\rm rot}$), but radio beams should shrink for pulsars having lower values of $\dot{E}_{\rm rot}$. The radio emission for high-$\dot{E}_{\rm rot}$ pulsars should therefore originate in wide beams at a significant fraction of $R_{\rm LC}$. One should however bear in mind that LCs resulting from radio and \g caustics would generally be nearly phase-aligned (although small phase differences could result if the radio and \g emission regions are at different altitudes). Caustic radio emission is therefore plausible for young pulsars with nearly aligned LCs. Radio caustics may however be more common in the case of the MSPs, as there are many more examples of MSPs with phase-aligned LCs.

\subsection{Modeling Phase-Aligned LCs}
We investigate the possibility of reproducing phase-aligned radio and \g LCs using ``altitude-limited'' OG / TPC models (alOG / alTPC) in which we limit the extent of the emission regions (Section~\ref{sec:alt-lim}) vs.\ a low-altitude SG (laSG) model (Section~\ref{sec:laSG}). 
By modeling the LCs of \JninthC \JfirstC and \JblackC we can infer values for the magnetic inclination and observer angles $\alpha$ and $\zeta$ (Section~\ref{sec:Results}), and also constrain the emission altitudes.

\section{BACKGROUND ON SELECTED MSPs WITH PHASE-ALIGNED LCs}
\label{sec:Background}

\subsection{\Jninth}
\Jninth was discovered using the Parkes radio telescope~\cite{Bailes94}. It follows a circular orbit around a low-mass companion (\textit{Hubble Space Telescope} observations revealed an optical white dwarf companion with a mass of about $0.2M_\odot$; \cite{Bell95}). \JninthN's period of $P=1.87$~ms implies a rotational age of $\tau_{\rm c}=P/2\dot{P}\sim10$~Gyr, dipolar surface field of $B_0\sim10^8$~G, and spin-down power of $\dot{E}_{\rm rot}\sim2\times10^{34}$~erg s$^{-1}$ \cite{Abdo09_J0034}, typical among the radio MSP population. It is also relatively close, lying at 0.5~kpc \cite{Hobbs05}. No rotating vector model (RVM) fits exist for \Jninth \cite{Stairs99} because there is no detected linear polarization. A $<3\sigma$ detection of an X-ray source 0.2$^{\prime\prime}$ from the pulsar position by \textit{XMM-Newton} was reported~\cite{Zavlin06}, while \textit{EGRET} obtained a $3\sigma$ flux upper limit above 100~MeV \cite{Fierro95} exceeding the recent \textit{Fermi} flux measurements \cite{Abdo09_J0034} by an order of magnitude.

\subsection{\Jfirst (B1937+21)}
\Jfirst is the first MSP ever discovered~\cite{Backer82}. It has a period $P=1.558$~ms, and remains one of the fastest-spinning MSPs discovered. This MSP has a very high spin-down luminosity of $\approx10^{36}$~erg s$^{-1}$, surface magnetic field of $B_0\sim4\times10^8$~G, and characteristic age of $\tau_{\rm c}\sim0.2$~Gyr~\cite{ATNF}, and lies at a distance of $d=7.7\pm3.8$~kpc \cite{Verbiest09,Guillemot11_2msps}. \textit{RXTE} observations \cite{Cusumano03} revealed a double-peaked X-ray profile with phase separation of about half a rotation, closely aligned with the phases of the giant radio pulses~\cite{Kinkhabwala00}, but slightly lagging the radio peaks. The \textit{EGRET} $3\sigma$ upper limit to the unpulsed flux above 100~MeV was $15.1\times10^{-8}$~cm~s$^{-1}$~\cite{Fierro95}. Pulsations with a significance well above $5\sigma$ have now been detected from \Jfirst by \textit{Fermi} LAT~\cite{Guillemot11_2msps}.

\subsection{\Jblack (B1957+20)}
\Jblack was the first ``black widow'' pulsar discovered. It is  in a nearly circular eclipsing binary orbit, ablating its low-mass tidally-locked companion star
~\cite{Fruchter88,VanParadijs88}. 
Lying at a distance of $d\sim2.5$~kpc, its period $P=1.607$~ms and intrinsic $\dot{P}$ is $\sim8\times10^{-21}$~\cite{Guillemot11_2msps} lead to a spin-down luminosity of $\sim7\times10^{34}$~erg s$^{-1}$, surface magnetic field of $B_0\sim10^8$~G, and characteristic age of $\tau_{\rm c}\sim3$~Gyr. \textit{XMM-Newton} observations revealed a phase dependence of the X-ray emission on the binary orbital period~\cite{Huang07}, although no pulsations were detected at the pulsar spin period $P$. A $\sim4\sigma$ pulsed X-ray signal have now been observed from \JblackC with the X-ray peaks seemingly in close alignment with the radio peaks \cite{Guillemot11_2msps}. Significant pulsed \g emission has also been detected by \textit{Fermi} LAT \cite{Guillemot11_2msps}.

\section{GEOMETRIC PULSAR MODELS}
\label{sec:Model}
As in in our previous work~\cite{Venter_MSP09}, we assumed a retarded vacuum dipole magnetic field as the basic structure of the pulsar magnetosphere \cite{Deutsch55,Dyks04_B}. In the case of young pulsars, this field may actually be closer to the force-free solution~\cite{Spitkovsky2006}, but it is not clear whether MSPs produce enough pairs to facilitate a force-free magnetosphere. 

We furthermore use a photon emission rate that is constant along magnetic field lines in the corotating frame, and treat relativistic effects (i.e., aberration of photon directions and time-of-flight delays; Section~\ref{sec:caustics}) consistently to first order in $r/R_{\rm LC}$ (with $r$ the radial distance from the NS center). We lastly include the Lorentz transformation (a second-order effect in $r/R_{\rm LC}$) of the local magnetic field between lab and corotating frames~\cite{BS10}.

\subsection{High-altitude Gamma-ray and Radio Emission: alOG and alTPC Models}
\label{sec:alt-lim}
\begin{figure}
\includegraphics[width=80mm]{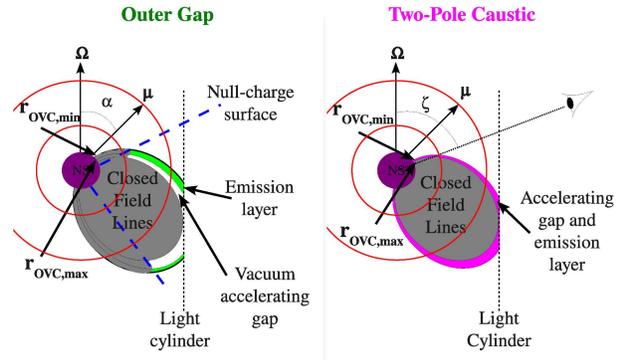}
\caption{Schematic diagram of the emission layers in the OG (panel~a) and TPC (panel~b) geometries. The magnetic axis is indicated by $\vec{\mu}$, and the spin axis by $\vec{\Omega}$. The two concentric circles indicate the limiting minimum and maximum emission radii imposed for the altitude-limited models.}
\label{fig:emission_layers}
\end{figure}

We use the same framework as previously~\cite{Venter_MSP09}, but the radio emission region is extended in altitude. We free the minimum and maximum radii of the radio ($R^r_{\rm min}$ and $R^r_{\rm max}$) and the maximum radius of the \g ($R^\gamma_{\rm max}$) emission regions, and restrict the emission gaps' extent to a cylindrical radius $\rho_{\rm max}<0.95R_{\rm LC}$. Importantly, we do not use an axis-centered radio conal model, but investigate radio photons coming from an OG / TPC-like structure. In the alOG radio models, the minimum radius is actually ${\rm max}\left\{R^r_{\rm min},R_{\rm NCS}\right\}$, so it is a function of magnetic azimuth $\phi$ and co-latitude $\theta$ when $R_{\rm NCS} > R^r_{\rm min}$. Here, $R_{\rm NCS}(\theta,\phi)$ is the radius of the NCS. We always set $R^\gamma_{\rm min}=R_{\rm NS}$ for alTPC (and TPC) and $R^\gamma_{\rm min} = R_{\rm NCS}$ for alOG (and alOG) fits, while $R^r_{\rm min}$ may vary and may even be quite close to $R_{\rm LC}$.  
Our alOG and alTPC models have~9 and~8 free parameters respectively, describing the pulsar geometry ($\alpha$ and $\zeta$) and \g and radio gap locations, apart from $P$ which determines the size of the PC (see Figure~\ref{fig:emission_layers}). More details are provided in~\cite{Venter11}. Note that the radio and \g emission layers are fit independently. 

\subsection{Low-altitude Gamma-ray and Radio Emission: laSG Models}
\label{sec:laSG}
This model provides a non-caustic explanation for the emission, and may be viewed as a geometric low-altitude SG model~\cite{MH03_SG} resembling a hollow cone beam close to the stellar surface. We modulate the emissivity according to
\begin{equation}
I \propto \left\{
    \begin{array}{ll}
      \exp\left(\Delta s/\sigma_{\rm in}\right),\quad s \leq s_{\rm f}\\
      \exp\left(-\Delta s/\sigma_{\rm out}\right),\quad s > s_{\rm f},\label{eq:PC}
    \end{array}
    \right.
\end{equation}
with $s$ the distance above the NS surface along a magnetic field line, $\Delta s \equiv s - s_{\rm f}$, and $\sigma_{\rm in}$ and $\sigma_{\rm out}$ setting the rate at which the intensity rises and falls along the magnetic field lines. The peak intensity occurs at a distance $s = s_{\rm f}$ (i.e., $\Delta s = 0$) along the field lines. The laSG models have~5 free parameters (more details in \cite{Venter11}).

\section{FINDING OPTIMAL LC FITS}
\label{sec:MCMC}
In order to statistically pick the best-fit parameters, for the alOG and alTPC models and for the three MSPs considered here, we have developed an Markov chain Monte Carlo (MCMC) maximum likelihood procedure~\cite{Johnson11}. The gamma-ray LCs are fit using Poisson likelihood while the radio LCs are fit using a $\chi^{2}$ statistic, and the two values are then combined. For a given parameter state the likelihood value is calculated by independently optimizing the radio and \g model normalizations using the \emph{scipy} python module\footnote{See http://docs.scipy.org/doc/ for documentation.} and the \emph{scipy.optimize.fmin\_l\_fbgs\_b} multivariate, bound optimizer~\cite{Zhu97}. 

An MCMC involves taking random steps in parameter space and accepting a step based on the likelihood ratio with respect to the previous step~\cite{H70}. The likelihood surfaces can be very multimodal which can lead to poor mixing of the chain and slow convergence. Therefore, we have implemented small-world chain steps \cite{G06} and simulated annealing \cite{MP92} to speed up the convergence and ensure that the MCMC fully explores the parameter space and does not get stuck in a local maximum. We verify that our chains have converged using the criteria proposed by \cite{GR92}.

In order to balance the gamma-ray and radio contributions to the likelihood, we have chosen to use a radio uncertainty which is equal to the average relative \g uncertainty in the on-peak region times the maximum radio value. The choice of radio uncertainty can strongly affect the best-fit results; in particular, a smaller uncertainty will decrease the overall likelihood and can, in some cases, lead to a different best-fit geometry which favors the radio LC more strongly. 
When varying the radio uncertainty by a factor of~2, the best-fit $\alpha$ and $\zeta$ values of \Jninth were found to change by $\lesssim13^{\circ}$.  For \JfirstC the best-fit $\alpha$ and $\zeta$ were found to vary by $\leq7^{\circ}$ when varying the radio uncertainty. The best-fit geometry of \Jblack was found to be the most sensitive to changes in the radio uncertainty, with either the best-fit $\alpha$ or $\zeta$ value changing by $\sim35^{\circ}$, while the other parameter changed by $\lesssim15^{\circ}$. 

Starting from the best-fit parameters found by the MCMC, we produced confidence contours in $\alpha$ and $\zeta$ by performing likelihood profile scans over the other parameters, allowing for the possibility of finding a better fit. The uncertainties on $\alpha$ and $\zeta$ quoted in Table~\ref{tab2} are approximate 95\% confidence level uncertainties.
We can also estimate uncertainties on the emission altitude parameters using the information from the likelihood profile scans which generated the confidence contours. 
Note that we used manually-selected LC fits for the laSG models. 

\section{RESULTS}
\label{sec:Results}
\begin{figure}
\includegraphics[width=80mm]{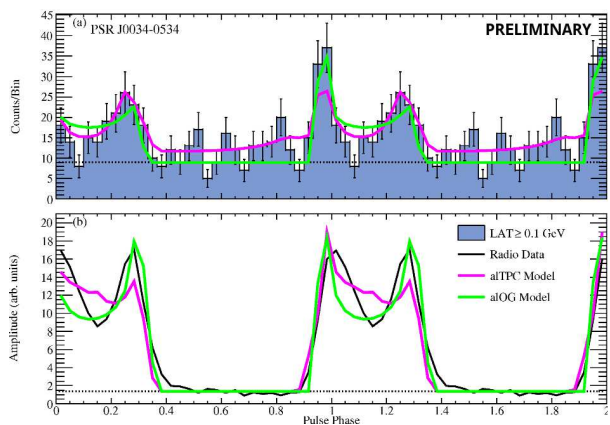}
\caption{LC fits for \Jninth using alOG / alTPC models. Panel~(a) shows the \g data, while panel~(b) shows the radio data.}
\label{fig:J0034_lim}
\end{figure}
\begin{figure}
\includegraphics[width=80mm]{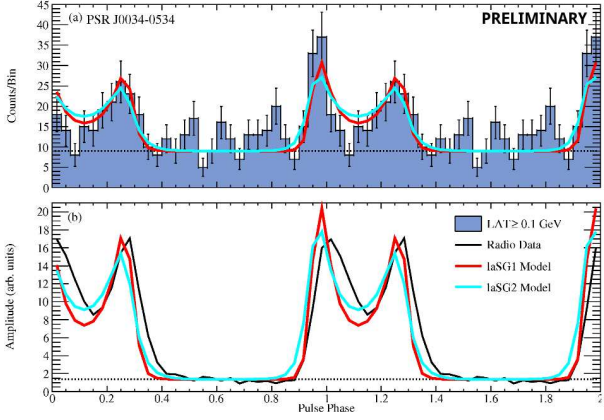}
\caption{LC fits for \Jninth using laSG models. Note that we had to introduce a large phase shift of $\phi_0=0.68$ for both \g and radio LCs, as the data and model zero phases do not coincide. This implies that the definition of `leading peak' and `first peak' do not coincide anymore. See Table~\ref{tab2} for more details.}
\label{fig:PC_J0034}
\end{figure}

The effects of letting $R_{\rm min}$ and $R_{\rm max}$ be free parameters in the alOG / alTPC model context, as well as using different fading parameters in our laSG models is discussed elsewhere~\cite{Venter11}. Our best-fit LC parameters are summarized in Table~\ref{tab2}. As an example, the alOG / alTPC LC fits for \Jninth are shown in Figure~\ref{fig:J0034_lim}, while Figure~\ref{fig:PC_J0034} shows fits for \Jninth in the case of laSG models. 

\section{CONCLUSIONS}
\label{sec:Con}
We studied a third subclass of \g MSP LCs for which the \g and radio profiles are phase-aligned, and the \g and radio emission should therefore be co-located. We introduced free parameters $R_{\rm min}$ and $R_{\rm max}$ in the alOG / alTPC models, (both are free for the radio emission region, but only $R^\gamma_{\rm max}$ is free for the the \g emission region) and found fits from these models which could reproduce the salient features of the profiles, although not perfectly. As a second option, we implemented the the laSG models and demonstrated that a modulated emissivity at low altitudes can reproduce main features of the profiles quite well. 

At the moment, it is difficult to quantitatively favor one class of models above the other, since searching for optimal laSG LC fits has been done manually. However, we did calculate the likelihood of the best-fit laSG LCs. The alTPC models provide slightly better LC fits than the alOG models, and both of these give better fits than the laSG models for the parameters listed (see Table~\ref{tab2}). Favoring the alOG / alTPC models over the laSG model therefore implies that the phase-aligned \g and radio LCs are most probably of caustic origin, produced in the outer magnetosphere, and the radio emission is most likely originating near the light cylinder. Thus, we can now divide the \g MSP population into three subclasses on the basis of their LCs: those with LCs fit by standard OG / TPC models, those with phase-aligned LCs fit by alOG / alTPC or laSG models, and those with LCs fit by PSPC models.

Radio polarization measurements can be used to give independent constraints on the pulsar viewing geometry \cite{Weltevrede10}, complementing the \g model fits, although the traditional RVM \cite{RC69} is not valid for the alTPC or alOG models where the radio peaks are caustics. Furthermore, the RVM is not expected to yield good results in the case of radio cone beam emission in MSPs, as these beams should suffer significant distortions due to retardation and aberration~\cite{BCW91}. This may account for the generally poor or non-existent RVM fits of MSP polarization data. 

Caustic models predict rapid PA swings coupled with depolarization~\cite{Dyks04_B}, since although the emission originates from a large range of altitudes and magnetic field orientations, it is restricted to a narrow phase interval to form the peaks. These features seem to be present in radio polarization measurements of the modeled MSPs~\cite{Stairs99,Yan11,Thorsett90}. Polarization signatures are important to help discriminate between models with caustic emission (such as occurs in alOG / alTPC models) and non-caustic emission (e.g., in the laSG model).

Future studies include development of full radiation models which will be able to reproduce the multiwavelength LC shapes, polarization properties, as well as the energy-dependent behavior of the spectra of the \g MSPs. Ways to increase pair production also need to be found, which may include investigation of offset-PC dipole magnetic fields~\cite{HM11} and higher-multipole magnetic fields near the NS surface~\cite{Zhang03}.

\begin{table}[b]
\begin{center}
\caption{Inferred best-fit model LC parameters for \JninthC \JfirstC and \JblackF The columns represent the geometric model (`laSG1' refers to an laSG model with $s_{\rm f}=1.2R$, $\sigma_{\rm in}=0.1R$, and $\sigma_{\rm out}=0.3R$, and `laSG2' refers to an laSG model with $s_{\rm f}=1.5R$, $\sigma_{\rm in}=0.2R$, and $\sigma_{\rm out}=0.5R$), inclination and observer angles $\alpha$ and $\zeta$ (measured in degrees), maximum \g altitude $R^\gamma_{\rm max}$, minimum radio altitude $R^r_{\rm min}$, maximum radio altitude $R^r_{\rm max}$, as well as the log-likelihood $\Lambda=-\Delta\ln({\rm like})$ of the fit. The altitudes are in units of $R_{\rm LC}$. We used $R^\gamma_{\rm min} = R_{\rm NCS}$ for the alOG model, and $R^\gamma_{\rm min} = R_{\rm NS}$ for the alTPC model. }
\vskip0.2cm
\begin{tabular}{|l|c|c|c|c|c|c|} 
\hline
Model & $\alpha$ & $\zeta$ & $R^\gamma_{\rm max}$ & $R^r_{\rm min}$ & $R^r_{\rm max}$ & $\Lambda$ \\
\hline
\hline
\multicolumn{7}{|l|}{\textbf{\Jninth}}   \\
\hline
alOG  & 12$^{+40}_{-6}$      & 69$^{+10}_{-2}$  & 0.9$^{+0.3}_{-0.1}$  & 0.2$^{+0.6}_{-0.06}$ & 1.1$^{+0.1}_{-0.4}$ & 96.1 \\
alTPC & 30$^{+9}_{-7}$      & 70$\pm2$      	& 0.9$\pm0.1$  & 0.7$^{+0.2}_{-0.3}$ & 0.8$^{+0.4}_{-0.1}$ &  87.0 \\
laSG1  & 10       & 34         & ---  & --- & --- & 97.3 \\
laSG2  & 10       & 37         & ---  & --- & --- & 98.7 \\
\hline
\multicolumn{7}{|l|}{\textbf{\Jfirst}}   \\
\hline
alOG  & 84$^{+2}_{-6}$      & 84$^{+1}_{-3}$      & 1.0$^{+0.2}_{-0.1}$ & 0.6$\pm0.1$ & 0.9$\pm0.1$ & 130.9 \\
alTPC & 75$^{+8}_{-6}$      & 80$^{+1}_{-3}$      & 1.0$\pm0.2$ & 0.7$^{+0.1}_{-0.3}$ & 0.9$^{+0.2}_{-0.1}$     &  126.3  \\
laSG1 & 30       & 32       & --- & --- & --- & 146.9 \\
laSG2 & 35       & 25       & --- & --- & --- & 154.6 \\
\hline
\multicolumn{7}{|l|}{\textbf{\Jblack}}   \\
\hline
alOG  & 31$^{+39}_{-3}$     & 89$^{+5}_{-3}$      & 1.1$^{+0.1}_{-0.2}$ & 0.7$\pm0.1$ & 0.9$^{+0.2}_{-0.1}$ & 128.3 \\
alTPC & 47$^{+5}_{-13}$     & 85$^{+1}_{-7}$      & 1.2$^{+0.1}_{-0.4}$ & 0.8$\pm0.1$ & 1.0$^{+0.2}_{-0.1}$     & 123.7 \\
laSG1 & 20     & 43       & ---   &  --- & --- & 129.0 \\
laSG2 & 25     & 45       & ---   &  --- & --- & 141.7 \\
\hline
\end{tabular}
\label{tab2}
\end{center}
\end{table}

\bigskip 
\begin{acknowledgments}
CV is supported by the South African National Research Foundation. AKH acknowledges support from the NASA Astrophysics Theory Program. CV, TJJ, and AKH acknowledge support from the \textit{Fermi} Guest Investigator Program as well as fruitful discussions with Dick Manchester and Matthew Kerr. Part of this work was performed at the Naval Research Laboratory and is sponsored by NASA DPR S-15633-Y.

The $Fermi$ LAT Collaboration acknowledges support from a number of agencies and institutes for both development and the operation of the LAT as well as scientific data analysis. These include NASA and DOE in the United States, CEA/Irfu and IN2P3/CNRS in France, ASI and INFN in Italy, MEXT, KEK, and JAXA in Japan, and the K.~A.~Wallenberg Foundation, the Swedish Research Council and the National Space Board in Sweden. Additional support from INAF in Italy and CNES in France for science analysis during the operations phase is also gratefully acknowledged.
\end{acknowledgments}

\end{document}